\documentclass{article}
\usepackage{graphicx} % Required for inserting images
\usepackage{bera}% optional: just to have a nice mono-spaced font
\usepackage{listings}
\usepackage{xcolor}
\usepackage[nonumberlist]{glossaries}
\usepackage{fancyhdr}
\usepackage{authblk}
\usepackage{orcidlink}
\usepackage{biblatex} %Imports biblatex package
\colorlet{punct}{red!60!black}
\definecolor{background}{HTML}{EEEEEE}
\definecolor{delim}{RGB}{20,105,176}
\colorlet{numb}{magenta!60!black}

\lstdefinelanguage{json}{
    basicstyle=\normalfont\ttfamily,
    numbers=left,
    numberstyle=\scriptsize,
    stepnumber=1,
    numbersep=8pt,
    showstringspaces=false,
    breaklines=true,
    frame=lines,
    backgroundcolor=\color{background},
    literate=
     *{0}{{{\color{numb}0}}}{1}
      {1}{{{\color{numb}1}}}{1}
      {2}{{{\color{numb}2}}}{1}
      {3}{{{\color{numb}3}}}{1}
      {4}{{{\color{numb}4}}}{1}
      {5}{{{\color{numb}5}}}{1}
      {6}{{{\color{numb}6}}}{1}
      {7}{{{\color{numb}7}}}{1}
      {8}{{{\color{numb}8}}}{1}
      {9}{{{\color{numb}9}}}{1}
      {:}{{{\color{punct}{:}}}}{1}
      {,}{{{\color{punct}{,}}}}{1}
      {\{}{{{\color{delim}{\{}}}}{1}
      {\}}{{{\color{delim}{\}}}}}{1}
      {[}{{{\color{delim}{[}}}}{1}
      {]}{{{\color{delim}{]}}}}{1},
}

\title{Prompt Engineering Guidance for Conceptual Agent-based Model Extraction using Large Language Models}
\author{
Siamak Khatami \orcidlink{0000-0003-4512-2661}, siamak.khatami@ntnu.no \\
Christopher Frantz \orcidlink{0000-0002-6105-8738}, christopher.frantz@ntnu.no}

\affil{
Norwegian University of Science and Technology, Department of Computer Science \\}
\date{Version 1.0 (November 2024)}
\makeglossaries
\newacronym{abm}{ABM}{Agent-based Modeling}
\newacronym{qa}{QA}{Question-answering}
\newacronym{llm}{LLM}{Large Language Model}
\newacronym{ai}{AI}{Artificial Intelligence}
\newacronym{nlp}{NLP}{Natural Language Processing}
\newacronym{json}{JSON}{JavaScript Object Notation}

\begin{document}

\maketitle
\begin{abstract}
    This document contains detailed information about the prompts used in the experimental process discussed in the article \emph{\textbf{\citetitle{ABMAutomationSCC}}}. The article aims to utilize \gls{qa} models to extract the necessary information to implement \gls{abm} from conceptual models. It presents the extracted information in formats that can be read by both humans and computers (i.e., \gls{json}), enabling manual use by humans and auto-code generation by \acrlong{llm}s (\acrshort{llm}).
\end{abstract}
% \printglossaries 
\textbf{Glossaries: }\acrfull{abm}, \gls{ai}, \acrfull{abm}, \gls{llm}, \gls{nlp}, \acrfull{qa}

\section{Introduction}

\acrfull{abm} is a simulation technique used to simulate complexities consisting of autonomic entities and heterogeneous systems to study and describe the complexity and causality of an event in the system \cite{gilbert2005simulation, Edmonds2017, GrimmABMIntro, WilenskyABM}. However, the development of \gls{abm} incorporates many development steps in which each step has its own characteristics, data types (qualitative and quantitative), and input/output \cite{Grimm2020}. This diversity leads to many development challenges, which have been discussed in \cite{ABMAutomationSCC}. These challenges were mainly caused by the diverse set of required skills (theoretical and computational), the iterative design, implementation, and evaluation refinement process, and extensive document sizes. One such important step is to implement simulation programs from conceptual models, which demands both field and programming knowledge. However, the emergence of \gls{ai} and especially \gls{nlp} has opened up hopes of facilitating these challenges.
\begin{figure}[t]
    \centering
    \includegraphics[width=\linewidth]{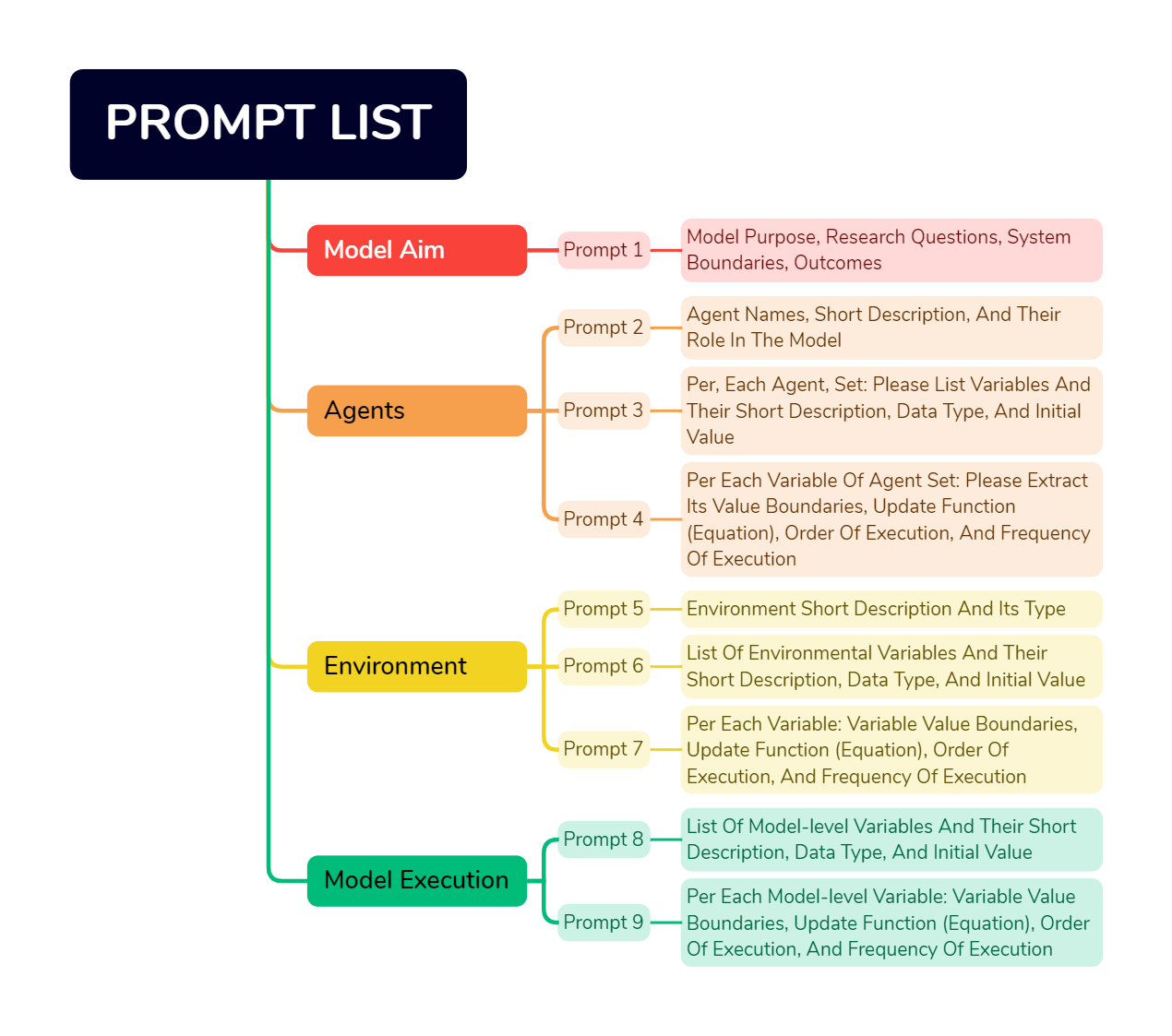}
    \caption{List of prompts and their target information}
    \label{fig:promptLists}
\end{figure}
In the Article \citetitle{ABMAutomationSCC}, the aim was to discuss the challenges presented by extracting information from conceptual models and implementing simulation model code. In this paper, by discussing different strategies to facilitate this problem (using auto-code generation models vs. \gls{qa} models), \gls{qa} models have been selected as a method to extract information from conceptual models. To develop a systematic approach relying on \acrshort{llm}s, the prompt engineering process, as well as desired input and output data structures, have been designed around distinctive target information commonly found in \gls{abm}s.

This document contains an overview of the final prompts used in that process. The aim of these prompts is to extract the required information described in Figure \ref{fig:promptLists} that highlights the prompts organized by core themes, namely modeling aim and purpose (`Model Aim'), agent information (`Agents'), environment information (`Environment') and finally information related to execution (`Model Execution')\footnote{For an extended discussion please consult the Article \cite{ABMAutomationSCC}.}.
% Requires: \usepackage{multirow}

\section{Model Aim}\label{Section:ModelAim}
The first step of any modeling technique is to clarify the aim and purpose. This step is crucial because it is not possible to create an exact replica of the world due to its complexities \cite{StermanJohnD.2000Bd:s}. Our goal is not to simulate an identical world but rather to understand it. If we fail to clarify our purpose, we may end up with another complex system that we can not understand \cite{StermanJohnD.2000Bd:s}. However, clarifying the modeling purpose is not limited to the mentioned reasons. All models are usually based on certain assumptions due to reasons like lack of information, abstraction from literature, and simplification \cite{Edmonds2017}. These assumptions and simplifications should be made in a way that does not undermine the main purpose of the modeling. Thus, it is beneficial to understand and clarify the model's purpose. 

\subsection{Prompt 1: Model purpose} \label{Prompt1}
This prompt set aims to extract a description of the model, relevant research questions, system boundaries, and independent and dependent variables.

All prompts follow the general structure of presenting the engine with an implicit role description, as well as clear instructions with respect to objectives, providing a partially populated desired output structure.

\begin{lstlisting}[language=json,firstnumber=1]
Analyze the provided ABM text to identify the purpose of the model, including a description, research questions, system boundaries, and outcome variables. Present the extracted data exclusively in JSON format, ensuring that the JSON object is comprehensive and contains all requested information. Avoid any form of data truncation or summarizing, and ensure that the response is strictly limited to the JSON object without any supplementary text. Please do not generate extra text and answer. The JSON should follow this structure:
{
    'Model_Purpose': {
        'full_description': Full_DESCRIPTION,
        'research_questions': [
            'RESEARCH_QUESTION_1',
            'RESEARCH_QUESTION_2',
            ... 
        ],
        'system_boundaries': [],
        'outcome_variables': {
            VAR1_NAME: SHORT_DESCRIPTION,
            VAR2_NAME: SHORT_DESCRIPTION,
            ....
            } 
    } 
} 
\end{lstlisting}

\section{Agent sets}\label{Section:AgentSets}
This series of prompts aims to extract information about the agent sets presented in the model and their relevant information. To ensure more consistent results, we avoided nested prompts, resulting in the current prompt structure shown in Prompts 2-4.

\subsection{Prompt 2}\label{Prompt2}
% List of Variables for Agent Sets, a short description, and their role
Prompt 2 is the first in a series of prompts (2, 3, and 4) related to agent sets in a conceptual document. This prompt aims to extract a list of agent sets within the conceptual model, along with a short description and an explanation of their roles in the model. The names of the agent sets will be used in the subsequent prompts (3 and 4) to explicitly extract more information relevant to each agent set.

\begin{lstlisting}[language=json,firstnumber=1]
Analyze the provided ABM text to identify the list of all agent sets, a short description, and their agent role in the system. Present the extracted data exclusively in JSON format, ensuring that the JSON object is comprehensive and contains all requested information. Avoid any form of data truncation or summarizing, and ensure that the response is strictly limited to the JSON object without any supplementary text. The JSON should follow this structure:

{
    AGENT_SET_1_NAME: { 
        'short_description':SHORT_DESCRIPTION, 
        'agent_role': SHORT_DESCRIPTION_AGENT_ROLE
    },
    ...
} 
\end{lstlisting}  
\subsection{Prompt 3}\label{Prompt3}

Prompt 3 focuses on extracting information about variables, including a brief description of each variable, its data type, and its initial value. As stated in Prompt 2 (Section \ref{Prompt2}), the agent set name must be extracted first. This prompt will then run for each agent set that has been extracted.

\begin{lstlisting}[language=json,firstnumber=1]
Input: {AGENT_SET_NAME}
\end{lstlisting}  

\begin{lstlisting}[language=json,firstnumber=1]
Analyze the provided ABM text to identify and extract the complete list of variables, variable data type, and initial value related to the '{AGENT_SET_NAME}' agent. Please ensure you extract all variables and
characteristics. Present the extracted data exclusively in JSON format, ensuring that the JSON object is comprehensive and contains all requested information. Avoid any form of data truncation or summarization, and ensure that the response is strictly limited to the JSON object without any supplementary text. The JSON should follow this structure:

{ 
    '{AGENT_SET}' :{
        VAR1: {
            'short_description': SHORT_DESCRIPTION,
            'data_type': DATA_TYPE, 
            'initial_value':INITIAL_VALUE,
        }, 
        VAR2 :{...} 
    } 
} 
\end{lstlisting}

\subsection{Prompt 4}

Prompt 4 is the last prompt in the series of prompts relevant to the agent-set section. Following the initial list of variables in prompt 3 relevant to the extracted agent sets using prompt 2, prompt 4 aims to extract information relevant to value boundaries, equations, frequency of execution, and order of execution for each variable. Prompt 4 can be merged with prompt 3 to reduce the quantity of prompts. However, experiments presented in \cite{ABMAutomationSCC} reported a lack of accuracy for the current \gls{llm}s. Thus, to avoid the nested and complex prompt, agent variable details have been separated into prompts 3 and 4. Agent set names (from Prompt 2) besides the specific variable name (from Prompt 3) are required to be imported in addition to the variable name. 

\begin{lstlisting}[language=json]
Inputs: '{AGENT_SET_NAME}', '{VAR}' from prompt 2 and 3
\end{lstlisting}  

\begin{lstlisting}[language=json,firstnumber=1]
Analyze the provided ABM text to identify and extract the value boundaries, equation, order of execution, and execution frequency related to the '{VAR}' variable of the '{AGENT_SET_NAME}' agent. Please ensure you extract all variables and characteristics. Present the extracted data exclusively in JSON format, ensuring that the JSON object is comprehensive and contains all requested information. Avoid any form of data truncation or summarization, and ensure that the response is strictly limited to the JSON object without any supplementary text. The JSON should follow this structure:

{ 
    '{AGENT_SET_NAME}': { 
        '{VAR}': {
            'value_boundaries':VALUE_BOUNDARIES,
            'equation': EQUATION,
            'order_number':ORDER_NUMBER, 
            'frequency': FREQUENCY
        }
    }
}
\end{lstlisting}  

\section{Environment, Space, and Grids}\label{Section:Environment}

Based on Figure \ref{fig:promptLists}, information relevant to the environment (also known as the simulation area, grid, ground, or cells) is another important part of the conceptual model. This aspect of simulation models typically includes auxiliary variable information that the thematician or programmer incorporates to visualize or address technical aspects of the model. The auxiliary variables are technical information that may not be presented in the real world or the theory of the thematician but are provided to facilitate the logic of the program. Therefore, extracting and verifying environment information is even more crucial than the agent sets information because it is subjective and relative within the context of the model.

In prompt sets 5, 6, and 7, we will focus on extracting information related to the environment (space) of the model presented in the conceptual framework. 

\subsection{Prompt 5}

This is the first prompt related to the environment of a model. In agent-based simulations, there are various types of environments, commonly referred to as grid types. To begin, it is important to identify the grid type and provide a brief description of it. While there is a possibility of exiting auxiliary information in addition to real environmental information (Section \ref{Section:Environment}), there is a chance of null information as well. Therefore, it is advisable to prevent the model from automatically generating non-existent facts. To manage this data augmentation, we utilize model instructions, which are outlined in Section \ref{Section:QAInstructions}. 

Instructions consist of a set of commands that are common to all prompts. This approach allows us to avoid repeating the same text in every prompt by having it fixed as instructions to the model.

\begin{lstlisting}[language=json,firstnumber=1]
Analyze the provided ABM text to identify and extract the information about the ABM simulation environment (space) type and environment (space) short description. Present the extracted data exclusively in JSON format, ensuring that the JSON object is comprehensive and contains all requested information. Avoid any form of data truncation or summarization, and ensure that the response is strictly limited to the JSON object without any supplementary text. The JSON should follow this structure:

{
    'Space': {
        'short_description':SHORT_DESCRIPTION,
        'type': TYPE 
    }
}
\end{lstlisting}

\subsection{Prompt 6}

Similar to the agent-set information in prompt 3, this prompt aims to extract a list of variables related to the environment. Each grid can contain cells, nodes, or connections, and each of these entities can have different characteristics and properties. For example, a cell in a 2D grid may have characteristics such as its Coordinates in the grid, background color code, etc. However, to enhance the accuracy of the \gls{llm}, we avoided using nested or lengthy prompts. Therefore, information related to the environments has been separated into two distinct prompts (6 and 7).
\begin{lstlisting}[language=json,firstnumber=1]
Analyze the provided ABM text to identify and extract the complete list of variables, variable short description, data type, and initial value related to the model space. Please ensure you extract all variables and characteristics. Present the extracted data exclusively in JSON format, ensuring that the JSON object is comprehensive and contains all requested information. Avoid any form of data truncation or summarization, and ensure that the response is strictly limited to the JSON object without any supplementary text. The JSON should follow this structure:

{
    'SPACE': {
        VAR1:{ 
            'short_description':SHORT_DESCRIPTION,
            'data_type': DATA_TYPE, 
            'initial_value':INITIAL_VALUE, 
        },
        VAR2 :{...}
    }
}
\end{lstlisting}   

\subsection{Prompt 7}

Following the extracted environment variable list using Prompt 6, using this prompt per each variable, we try to extract information relevant to value boundaries, equations, the frequency of execution, and execution order. 

\begin{lstlisting}[language=json,firstnumber=1]
Input: '{VAR}' from prompt 6
\end{lstlisting}  

\begin{lstlisting}[language=json,firstnumber=1]
Analyze the provided ABM text to identify and extract the value boundaries, equation, execution order, and execution frequency related to the '{VAR}' variable of model space. Please ensure you extract all variables and characteristics. Present the extracted data exclusively in JSON format, ensuring that the JSON object is comprehensive and contains all requested information. Avoid any form of data truncation or summarization, and ensure that the response is strictly limited to the JSON object without any supplementary text. The JSON should follow this structure:

{
    'SPACE': {
        '{VAR}': {
            'value_boundaries': VALUE_BOUNDARIES,
            'equation': EQUATION,
            'excution_order':EXCUTION_ORDER,
            'frequency': FREQUENCY
        }
    } 
}
\end{lstlisting}  

\section{Model Execution} \label{Section:ModelLevel}

In addition to agent sets and environments, there are a set of variables that are not necessarily related to the theory of the conceptual model but to facilitate the modeling process, like model parameters or report variables. Thus, similar auxiliary and subjective characteristics exist at this level of the model as well. Thus, it is beneficial to extract model-level information, considering its crucial role. Thus, using the next two prompts, we will try to extract information relevant to the model.
\subsection{Prompt 8}\label{Prompt 8}
Prompt one already aims to extract information relevant to the model's aim and purpose; in this section, we will focus on the more technical side of the conceptual model and try to extract the variable list plus its short description, data types, and initial values. The remaining information will be extracted using Prompt 9 per each variable. 

\begin{lstlisting}[language=json,firstnumber=1]
Analyze the provided ABM text to identify and extract the complete list of model-level variables, variable data type, and initial value. Please ensure you extract all variables and characteristics only at the model level. Present the extracted data exclusively in JSON format, ensuring the JSON object is comprehensive and contains all requested information. Avoid any form of data truncation or summarization, and ensure that the response is strictly limited to the JSON object without any supplementary text. The JSON should follow this structure:

{
    'Model-Level': {
        VAR1: {
            'short_description':SHORT_DESCRIPTION, 
            'data_type': DATA_TYPE, 
            'initial_value':INITIAL_VALUE, 
        }, 
        VAR2 :{...} 
    } 
}
\end{lstlisting}  

\subsection{Prompt 9}\label{Prompt9}
Similar to agent sets and environment variables, using this prompt (9), we aim to extract more relevant information for each variable at the model level.

\begin{lstlisting}[language=json,firstnumber=1]
Input: '{VAR}' from Prompt 8
\end{lstlisting}  
\begin{lstlisting}[language=json,firstnumber=1]
Analyze the provided ABM text to identify and extract the value boundaries, equation, execution order, and execution frequency related to the '{VAR}' variable of Model-level variables. Please ensure you extract all variables and characteristics. Present the extracted data exclusively in JSON format, ensuring the JSON object is comprehensive and contains all requested information. Avoid any form of data truncation or summarization, and ensure that the response is strictly limited to the JSON object without any supplementary text. The JSON should follow this structure:

{
    'Model-Level': {
        '{VAR}': {
            'value_boundaries': VALUE_BOUNDARIES,
            'equation': EQUATION, 
            'order_number':ORDER_NUMBER, 
            'frequency': FREQUENCY
        }
    }
}
\end{lstlisting}  

\section{QA Model Instructions}\label{Section:QAInstructions}
Any QA model can have three types of inputs: instructions, prompts, and attachments, which each play a different role. Instructions are a set of commands that we try to include in all prompts; thus, instead of including the same repetitive command in every prompt, we instruct the \gls{qa} model to follow these instructions. For example, in \cite{ABMAutomationSCC}, the following instruction has been used in all prompts. 

\begin{lstlisting}[language=json,firstnumber=1]
You are an Agent-based modeling specialist. Your duty is to help users in extracting information from ABM texts for coding purposes. Get the user messages to extract the relevant information from the uploaded file. Do not summarize information; neither truncate nor auto-generate and augment any data if data and information are not presented in the provided document. Just return a full report of the expected information in the JSON format without any extra text around it. 
\end{lstlisting}

\printbibliography 
\end{document}